\documentclass[namedreferences]{kluwer}

\usepackage{epsfig}          

\begin{document}
\begin{article}

\begin{opening}



\title{Modeling gas and stellar kinematics in disc galaxies.}
\subtitle{NGC 772, NGC 3898, and NGC 7782}

\author{E. \surname{Pignatelli}\email{\tt pignatel@sissa.it}}
\institute{SISSA, via Beirut 2-4, I-34013 Trieste, Italy }
\author{J.C. \surname{Vega Beltr\'an}}
\author{J.E. \surname{Beckman}}
\institute{Instituto Astrof\'\i sico de Canarias, La Laguna, Spain }
\author{E.M. \surname{Corsini}}
\author{A. \surname{Pizzella}}
\institute{ Osservatorio Astrofisico di Asiago, Asiago, Italy}
\author{C.  \surname{Scarlata}}
\author{F.  \surname{Bertola}}
\institute{Dipartimento di Astronomia, Universit\`a di Padova,
   Padova, Italy }
\author{J.G.  \surname{Funes}}
\institute{Vatican Observatory, University of Arizona,
         Tucson, USA} 
\author{W.W.  \surname{Zeilinger}}
\institute{Institut f{\"u}r Astronomie, Universit{\"a}t Wien,
         Wien, Austria}



\end{opening}


\vspace*{-0.5cm}
We present $V-$band surface photometry and major-axis
kinematics of stars and ionized gas of three early-type
spiral galaxies, namely NGC 772, NGC 3898 and NGC 7782.
For each galaxy we built a self-consistent Jeans model
for the stellar kinematics, adopting the light
distribution of bulge and disc derived by means of a
two-dimensional parametric photometric decomposition.
This allowed us to investigate the presence of
non-circular gas motions, and derive the mass
distribution of luminous and dark matter in these
objects. We found that the observed gas rotation
corresponds to the circular velocity except for the
innermost region ($|r|\lsim 8''$) of NGC 3898. This
behaviour is quite common, although not ubiquitous, in
the few bulge-dominated galaxies, for which dynamical
modeling allows the comparison between the gas velocity
and the circular speed.  The mass is essentially traced
by light in NGC 772 and NGC 7782, where gas rotation
velocities were observed out to $0.2\,R_{25}$ and
$0.6\,R_{25}$, respectively. For NGC 3898 we succeed in
reproducing the observed gas rotation velocities, which
extends out to $0.7\,R_{25}$, only taking into account
the presence of a massive dark halo.

\vspace*{-0.2cm}
\begin{figure}[ht]
\vspace*{2.92cm}
\includegraphics{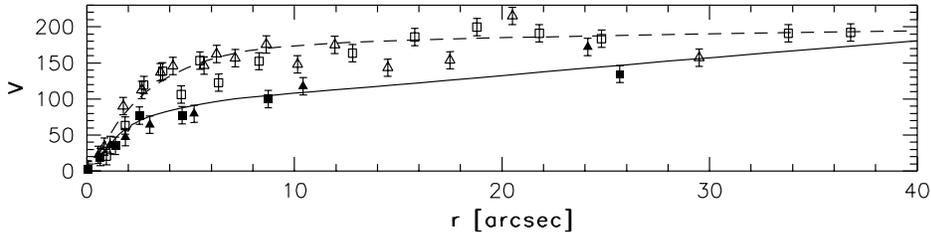}  
\caption{Gas ({\em open symbols}) and stars ({\em full symbols})
rotational velocities for NGC 772, against the model
circular ({\em solid line}) and rotational ({\em dotted
line}) velocities.}
\label{fig:v0772}
\end{figure}                                             


\end{article}
\end{document}